\documentclass[11pt,preprint]{aastex}
\usepackage{rotating}

\usepackage{lineno}

\usepackage{amsmath, amsthm, amssymb}
\usepackage{subfigure}
\usepackage{graphicx}

\usepackage{lineno}

\slugcomment{Submitted to ApJ}

\newcommand{\fermi} {\emph{Fermi }} 
\newcommand{\xmm} {\emph{XMM-Newton }}

\newcommand{\rosat} {\emph{ROSAT HRI}}

\def\hst{{\it Hubble Space Telescope }}

\newcommand{\be}{\begin{equation}}
\newcommand{\ee}{\end{equation}}

\begin{document}

\shorttitle{\emph{FERMI} OBSERVATIONS OF PSR~J1836+5925}

\shortauthors{\textsc{The Fermi LAT collaboration}}

\title{{\fermi} Large Area Telescope observations of PSR~J1836+5925}
\author{
A.~A.~Abdo\altaffilmark{2,3}, 
M.~Ackermann\altaffilmark{4}, 
M.~Ajello\altaffilmark{4}, 
W.~B.~Atwood\altaffilmark{5}, 
L.~Baldini\altaffilmark{6}, 
J.~Ballet\altaffilmark{7}, 
G.~Barbiellini\altaffilmark{8,9}, 
M.~G.~Baring\altaffilmark{10}, 
D.~Bastieri\altaffilmark{11,12}, 
K.~Bechtol\altaffilmark{4}
% ...
}
% ...
\altaffiltext{2}{Space Science Division, Naval Research Laboratory, Washington, DC 20375, USA}
\altaffiltext{3}{National Research Council Research Associate, National Academy of Sciences, Washington, DC 20001, USA}
\altaffiltext{4}{W. W. Hansen Experimental Physics Laboratory, Kavli Institute for Particle Astrophysics and Cosmology, Department of Physics and SLAC National Accelerator Laboratory, Stanford University, Stanford, CA 94305, USA}
\altaffiltext{5}{Santa Cruz Institute for Particle Physics, Department of Physics and Department of Astronomy and Astrophysics, University of California at Santa Cruz, Santa Cruz, CA 95064, USA}
\altaffiltext{6}{Istituto Nazionale di Fisica Nucleare, Sezione di Pisa, I-56127 Pisa, Italy}
\altaffiltext{7}{Laboratoire AIM, CEA-IRFU/CNRS/Universit\'e Paris Diderot, Service d'Astrophysique, CEA Saclay, 91191 Gif sur Yvette, France}
\altaffiltext{8}{Istituto Nazionale di Fisica Nucleare, Sezione di Trieste, I-34127 Trieste, Italy}
\altaffiltext{9}{Dipartimento di Fisica, Universit\`a di Trieste, I-34127 Trieste, Italy}
\altaffiltext{10}{Rice University, Department of Physics and Astronomy, MS-108, P. O. Box 1892, Houston, TX 77251, USA}
\altaffiltext{11}{Istituto Nazionale di Fisica Nucleare, Sezione di Padova, I-35131 Padova, Italy}
\altaffiltext{12}{Dipartimento di Fisica ``G. Galilei", Universit\`a di Padova, I-35131 Padova, Italy}
% ...

\begin{abstract}
The discovery of the $\gamma$-ray pulsar PSR~J1836+5925, powering the formerly unidentified EGRET source 3EG~J1835+5918, 
was one of the early accomplishments of the \emph{Fermi} Large Area Telescope (LAT). Sitting $25^\circ$ off the Galactic plane, PSR~J1836+5925 
is a 173\,ms pulsar with a characteristic age of 1.8 million years, a spindown luminosity of 1.1$\times10^{34}$\,erg s$^{-1}$, and a large 
off-peak emission component, making it quite unusual among the known $\gamma$-ray pulsar population. We present an analysis of one year of LAT data, 
including an updated timing solution, detailed spectral results and a long-term light curve showing no indication of variability. 
No evidence for a surrounding pulsar wind nebula is seen and the spectral characteristics of the off-peak emission indicate it is likely magnetospheric. 
Analysis of recent \xmm observations of the X-ray counterpart yields a detailed characterization of its spectrum, which, like Geminga, is 
consistent with that of a neutron star showing evidence for both magnetospheric and thermal emission.

\end{abstract}


\keywords{gamma rays: general; pulsars: general; pulsars: individual (PSR J1836+5925)}

\section{Introduction}
Since its discovery by EGRET~\citep{lin92}, the bright high-energy $\gamma$-ray source GRO~J1837+59 defied straightforward identification. It was 
reported as a persistent source with a varying flux of 3--8$\times10^{-7}$\,ph cm$^{-2}$ s$^{-1}$ and a relatively hard spectrum of photon index 1.7 
in non-consecutive, typically 2--3 week-long, observing periods. Its location 
at high Galactic latitude in conjunction with early reports of $\gamma$-ray 
variability~\citep[later questioned by][]{nolan96,reimer01} suggested it might be a blazar. However, the 
lack of a radio-bright counterpart, common to EGRET-detected blazars, cast doubts on such an interpretation.

With the detection of faint X-ray counterpart candidates in the error contour of 3EG~J1835+5918~\citep{reimer00}, the interpretation focused 
increasingly on a nearby radio-quiet neutron star. The complete characterization of all but one of the \rosat~X-ray sources was presented 
by \cite{mirabal00} and \cite{reimer01}, who singled out RX~J1836.2+5925 as the most probable counterpart of 3EG~J1835+5918. Subaru/FOCAS observations in 
the B- and U-bands proposed possible optical counterparts~\citep{totani02}, while \hst observations 
set an optical upper limit of $V > 28.5$~\citep{halpern02}. A scenario of a thermally emitting neutron star which was either older or more distant than 
the archetypal radio-quiet $\gamma$-ray pulsar Geminga emerged as the most plausible explanation for the source~\citep{halpern93,bignami96,mirabal01}, 
with an upper limit on the distance of 800\,pc, determined from X-ray observations~\citep{halpern02}. Using {\it Chandra} observations separated by 
three years, \cite{halpern07} were also able to determine an upper limit on the proper motion of 0.14\arcsec  per year, or $v_t < 530$ km s$^{-1}$ at 800\,pc.
However, a timing signature, which would settle the nature of this source, was never found in the EGRET data \citep{chandler01, ziegler08}, nor in 
repeated observations by {\it Chandra} \citep{halpern02, halpern07}, nor in a 24-hr observation with NRAO's Green Bank Telescope (GBT)~\citep{halpern07}. 
Upper limits from Very High Energy (VHE) $\gamma$-ray observations~\citep{fegan05} determined that the peak of emission must be at GeV gamma rays. 
Recently, AGILE reported marginal flux variability in their 2007--2008 data, arising from several non-detections in a 
period of long uninterrupted coverage, along with a flux level significantly lower than what was previously reported by EGRET~\citep{bulgarelli08}.

3EG J1835+5918 was a target of pointed observations during the 60-day commissioning period prior to the start of normal science operations of 
the Large Area Telescope (LAT) aboard the {\it Fermi Gamma-ray Space Telescope}. During these observations, the LAT accumulated photons from this source 
at a rate approximately twice as high as during regular survey-mode operations, thus facilitating the detection of $\gamma$-ray pulsations. 
The discovery and initial timing of the pulsar, PSR~J1836+5925, using the first five months of LAT data, were reported in \cite{BSPI}. Here we 
present the phenomenology emerging from one year of LAT observations of PSR~J1836+5925, including energy-dependent pulse profiles and
phase-resolved spectroscopy. 

\section{Gamma-ray observations and data analysis}

The LAT is a pair conversion telescope, sensitive to gamma rays with energies from 20\,MeV to $>$300\,GeV. Gamma rays in the LAT are 
recorded with an accuracy of $< 1\,{\mu}s$. The LAT has an on-axis effective area of 8000\,cm$^{2}$, a field of view of $\sim$2.4\,sr, and an 
angular resolution of $\sim$0.8$^\circ$ 68\% containment at 1\,GeV~\citep{atwood09}. 

\subsection{Timing analysis}

We have derived a precise timing solution of PSR~J1836+5925 using data from 2008 June 30 to 2009 June 30 (MJD 54647.4--55013.0). We selected 
photons with $E>170$ MeV offset from the source direction by no more than 1.6$^\circ$, a radius chosen to maximize the pulsed significance, and 
used the LAT Science Tool\footnote{available at http://fermi.gsfc.nasa.gov/ssc/data/analysis/} \texttt{gtbary} in its geocenter mode to 
correct the arrival times to terrestrial time (TT) at the geocenter. 
We generated a total of 22 pulse times of arrival (TOAs), each covering roughly two weeks of data, and obtained pulse profiles by folding the 
photon times according to a provisional ephemeris using polynomial coefficients generated by \textsc{Tempo2}~\citep{tempo2} in its predictive 
mode (assuming a fictitious observatory at the geocenter). The TOAs were then measured by cross correlating each pulse profile with a template 
consisting of two gaussians, derived from the data set above~\citep{ray10}. The timing model, fit using \textsc{Tempo2}, included position, 
frequency ($\nu$), and frequency derivative ($\dot{\nu}$). Table~\ref{timing_results} lists the results of our fit. The 1.3\,ms RMS residual to 
the fit is comparable to the mean TOA measurement uncertainty of 1.2\,ms and significantly smaller than the 5.4\,ms resolution of our 32 bin light 
curve. The reduced $\chi^2$ of our timing fit is 1.9.
With $\nu = 5.77$\,Hz and $\dot \nu = -5\times10^{-14}$\,Hz\,s$^{-1}$, we derive a characteristic age of 1.8 million years and a spindown luminosity of
$10^{34}$\,erg\,s$^{-1}$. Our best fit location is RA=18:36:13.75(3), Dec=+59:25:30.3(6), which is 0.35\arcsec\ from RX J1836.2+5925, well within the statistical 
uncertainty of the timing fit, securing the association between PSR~J1836+5925 and RX J1836.2+5925. 

\subsection{Light curve\label{lc}}

\begin{figure*}
\centering
\includegraphics[width=.75\columnwidth,trim=0 5 0 10]{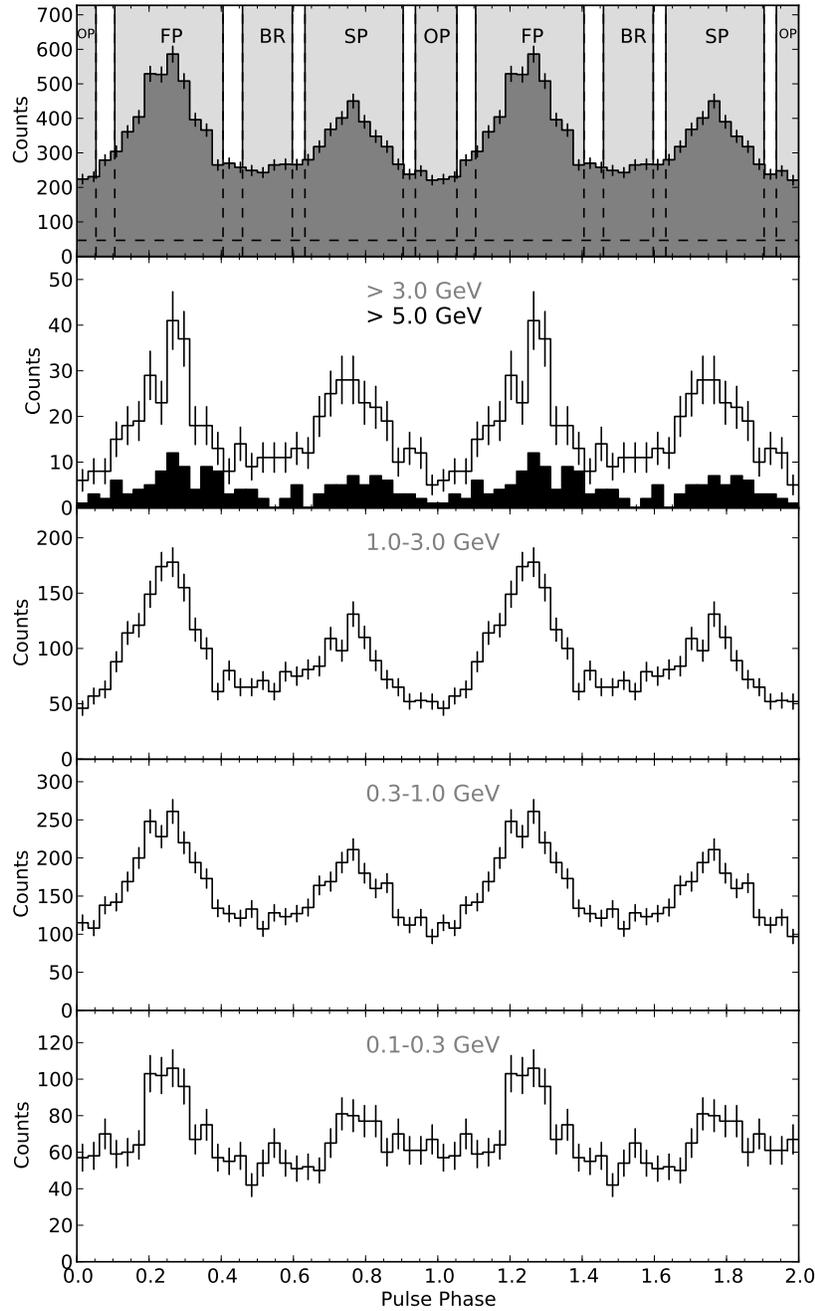}\\
\caption{Folded light curves of PSR~J1836+5925 with a resolution of 32 phase bins per period. Two rotations are shown. 
The top panel shows all events $>$100 MeV, along with the different phase regions labelled: 
Off-peak (OP), First Peak (FP), Bridge (BR), and Second Peak (SP) regions. The horizontal dashed line represents an estimate of the 
background due to diffuse emission, illustrating the high level of off-peak emission being emitted by the source. The lower four panels 
show the light curves in different energy bands. The darker histogram on the second panel from the top shows events with $E>5$\,GeV.}
\label{lightcurve}
\end{figure*}

We explored the pulsar light curve in different energy bands by selecting events with energies above 100 MeV from an energy-dependent 
region of interest (ROI), defined as $\theta=3.4^\circ(E/100 \mathrm{MeV})^{-0.75}$ with a minimum (maximum) radius of 0.35$^\circ$ (2.1$^\circ$). 
The rotation phase of each event is calculated using the truncated Taylor series expansion: $\phi=\phi_0+\nu(t-T_0)+\frac{1}{2}\dot\nu(t-T_0)^2$, where 
$T_0$ is the reference epoch of MJD 54800 and $\phi_0$ is the reference phase at $T_0$, which we define as $\phi_0$=0.55. 
Figure~\ref{lightcurve} (top panel) shows the folded light curve of the pulsar for energies above 100 MeV. The light curve has two distinct peaks 
and is well fit by a constant plus two gaussians centered at phases $\phi=0.26$ and $\phi=0.77$, with their means separated by $0.51\pm0.01$ in phase. 
The ``pulsed fraction" (determined by integrating the contribution from the two gaussians) is energy dependent, as can be appreciated from 
Figure~\ref{lightcurve}. It has an unusually low value of $\sim26 \pm 2$ \% for energies above 100 MeV, with the remaining $>$70\% coming from the 
constant term. While the pulsed fraction does increase at higher energies, up to $\sim$45\% above 1.5\,GeV, it is always less than 50\%, regardless 
of the cuts chosen.

We identify the following intervals: first peak (FP): 0.105$<\phi<$0.405, second peak (SP): 0.632$<\phi<$0.904, bridge (BR): 0.459$<\phi<$0.597, and off-peak (OP): 
0.938$<\phi<$0.053. The lower panels show the folded light curve in 5 different energy intervals: 0.1--0.3\,GeV, 0.3--1\,GeV, 1--3\,GeV, $>$3\,GeV, 
and $>5$\,GeV (dark histogram in the second panel). 

\subsection{Spectral Analysis}

We performed the spectral analysis using data collected during the sky survey: 2008 August 4 to 2009 June 30 (MJD 54682--55013). While 
the LAT data taken during the commissioning period are adequate for timing analyses, several of the configuration settings may have had a modest 
effect on the energy resolution and reconstruction, so we exclude these data for the spectral analysis. We also  
exclude events with zenith angles greater than $105^\circ$ to minimize the contamination from gamma rays from the Earth's atmosphere. A phase-averaged 
spectrum was obtained with an unbinned maximum likelihood analysis, using the LAT Science Tool \texttt{gtlike}~\citep{LATVela} and the ``Pass 6 v3" 
instrument response functions (IRFs). We used energies $>200$\,MeV. The diffuse emission from the Milky Way was modeled using 
\texttt{gll\_iem\_v02}$^1$ while the isotropic extragalactic diffuse emission 
and residual instrumental particle backgrounds were modeled together using the currently recommended ``template spectrum" 
\texttt{isotropic\_iem\_v02}$^1$. We extracted photons from a $15^{\circ}$ radius ROI centered on the coordinates of RX~J1836.2+5925, in order to properly account for 
the contributions of other gamma-ray sources in the vicinity. All sources from the LAT first-year source catalog~\citep{LATCatalog} included in our ROI were fit with a simple 
power law, while for the pulsar itself we used a power law with an exponential cutoff:

\begin{equation}
\frac{{\rm d} N}{{\rm d} E} = K E_{\rm GeV}^{-\Gamma}
                              \exp \left[- (\frac{E}{E_{\rm cutoff}})^n \right]
\label{expcutoff}
\end{equation}
where $\Gamma$ is the photon index, $E_{\rm cutoff}$ the cutoff energy, and $K$ the normalization (in units of ph cm$^{-2}$ s$^{-1}$ MeV$^{-1}$). A super-exponential cutoff ($n=2$) was ruled out, compared to a simple exponential ($n=1$) cutoff, at $8\sigma$ significance so we set $n$ equal to 1 in further analysis. Note that physical motivation for the use of a power-law spectrum with an exponential cutoff is underpinned by this being approximately the form expected for curvature or synchrotron radiation from both monoenergetic electrons \citep[e.g. see Eq. (24) of][]{harding08}, and electrons with a distribution of Lorentz factors up to some maximum value. 
A superexponential cutoff, on the other hand, would be expected in polar cap models, due to single photon pair production attenuation in strong magnetic fields near the surface~\citep{nel95,dh96,rh07}. The results of our spectral fits are summarized in Table~\ref{spectral_results}. The quoted errors are statistical only. The effect of the systematic uncertainties in the effective area on the spectral parameters is $\delta\Gamma$ = (+0.3, --0.1), $\delta E_{\rm cutoff}$ = (+20\%, --10\%), 
$\delta$F$_{100}$ = (+30\%, --10\%), and $\delta$G$_{100}$ = (+20\%, --10\%)~\citep{pulsarcatalog}.

\begin{figure*}[htp]
\centering
\subfigure[Off-peak (OP)]
{\label{OP}
\includegraphics[width=2.5in,angle=0]{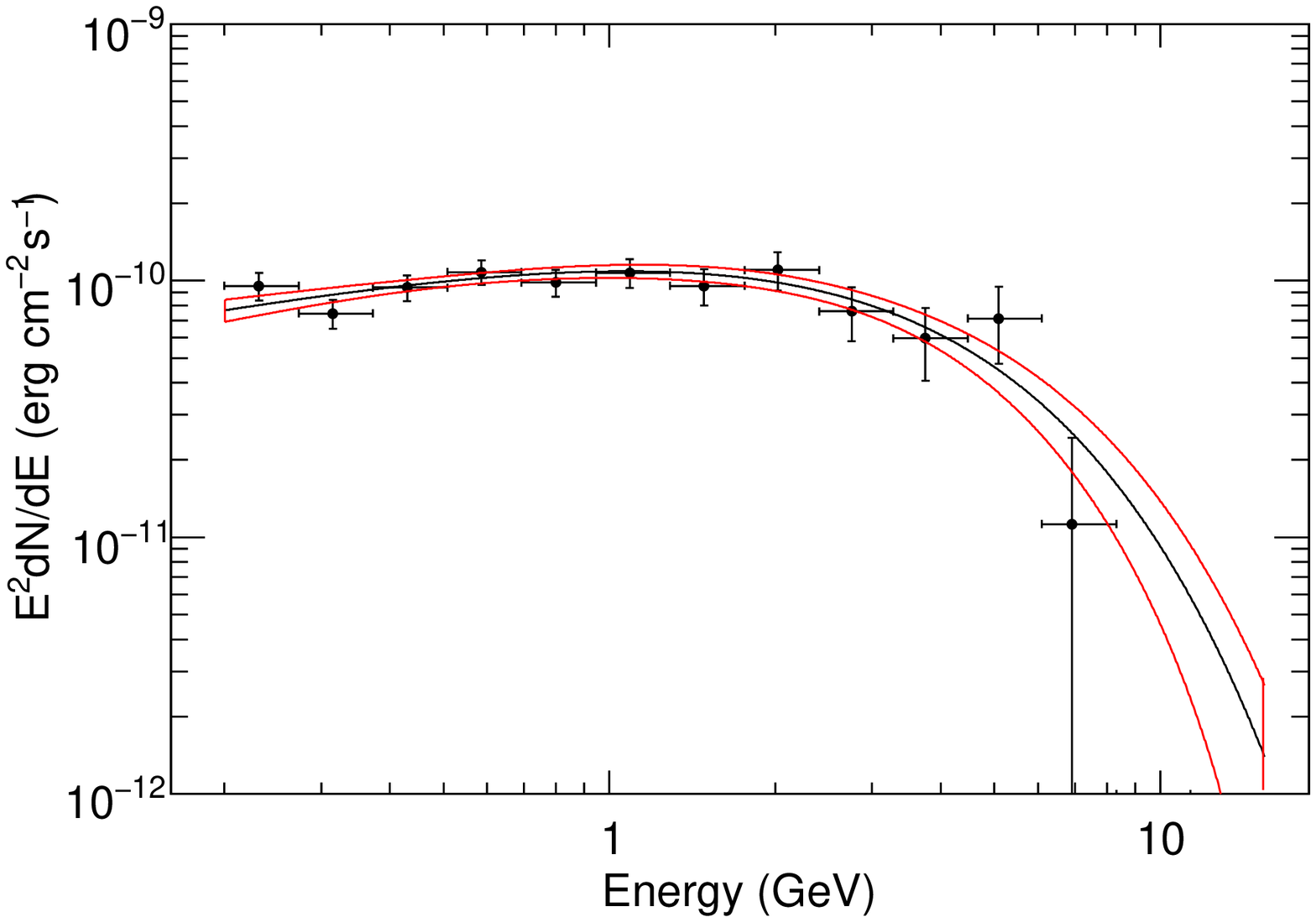}}
\subfigure[First Peak (FP)]
{\label{P1}
\includegraphics[width=2.5in,angle=0]{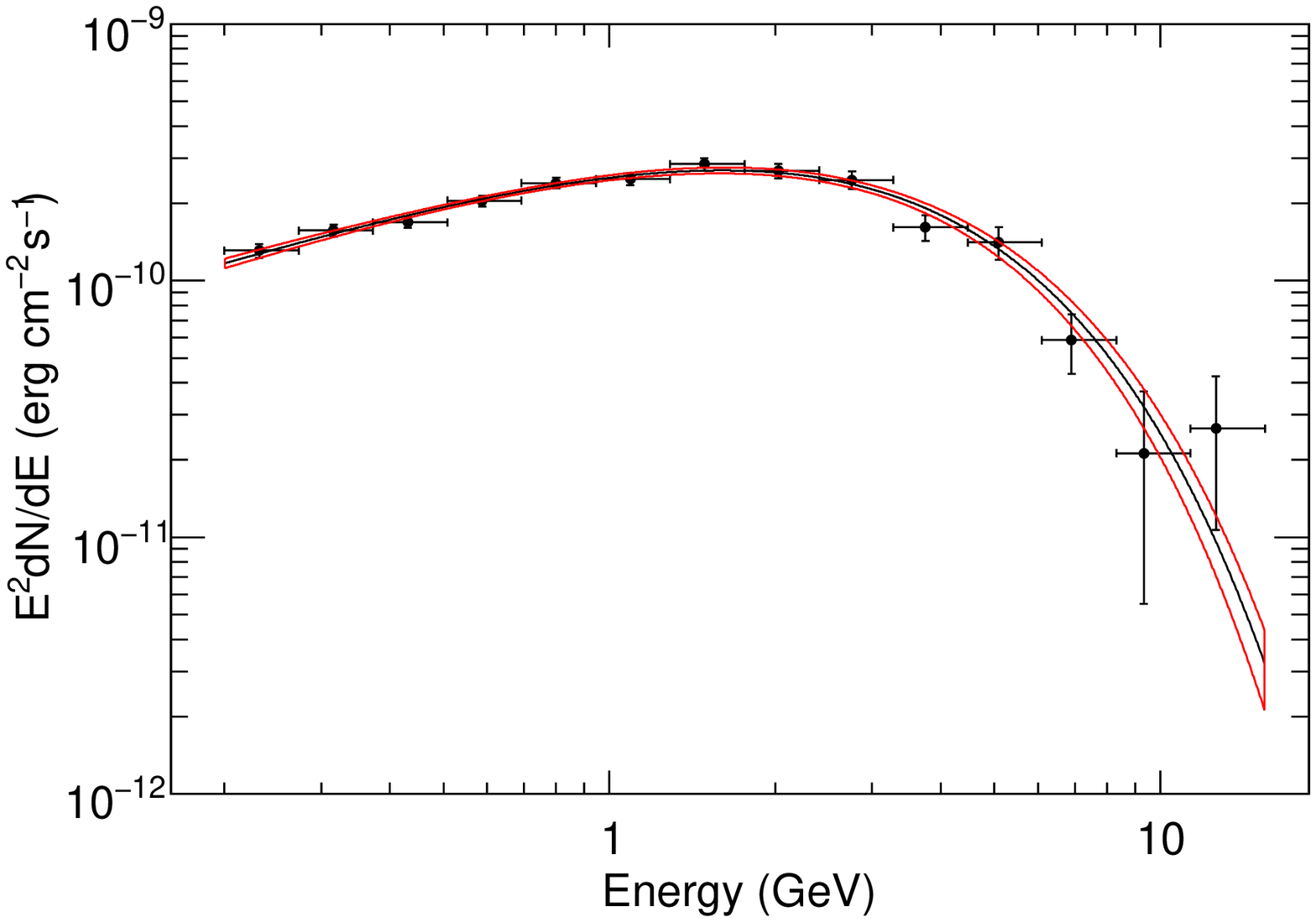}}
\subfigure[Bridge (BR)]
{\label{BR}
\includegraphics[width=2.5in,angle=0]{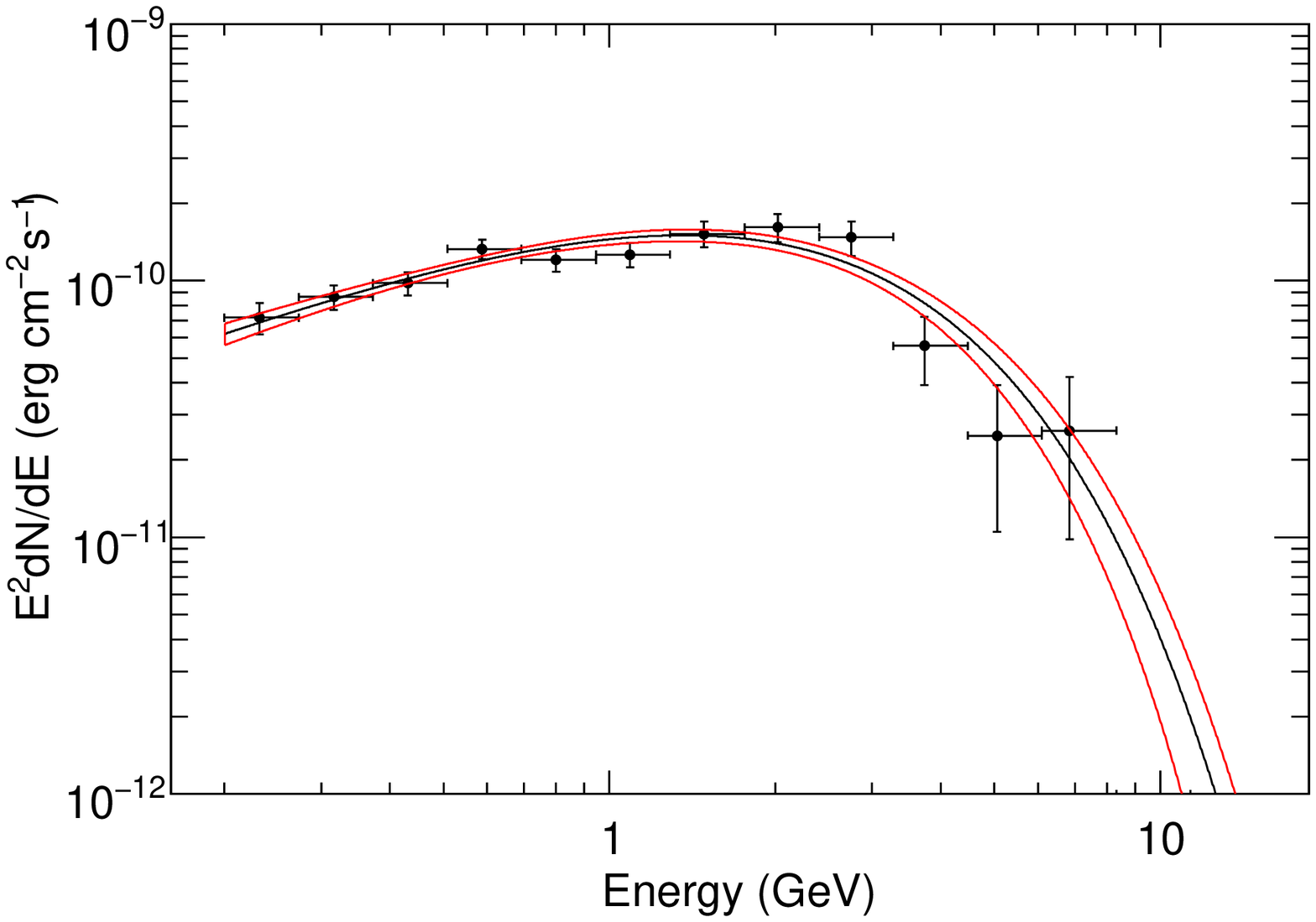}}
\subfigure[Second Peak (SP)]
{\label{SP}
\includegraphics[width=2.5in,angle=0]{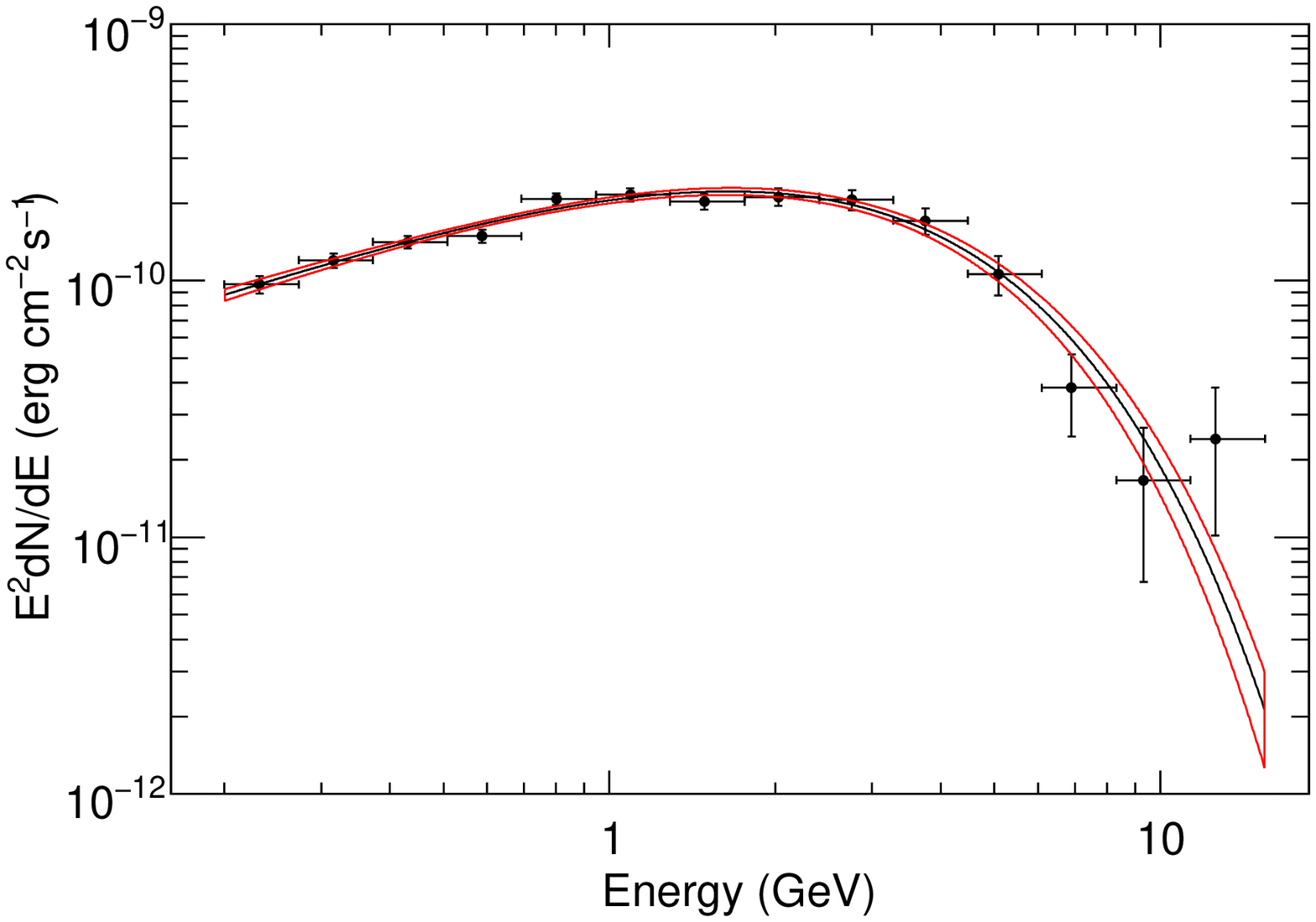}}
\caption{Energy spectra of the four identified phase regions of PSR~J1836+5925. The data points represent the measured fluxes obtained from likelihood 
fits in different representative energy bands where the pulsar is modeled as a power law, while the line shows the best-fit model obtained in the unbinned 
maximum likelihood analysis over the entire energy range, along with the 1$\sigma$ ``bowtie" confidence region. {\bf Top Left -- } Off-peak (OP). 
{\bf Top Right -- } First Peak (FP). {\bf Bottom Left -- } Bridge (BR). {\bf Bottom Right -- } Second Peak (SP).} \label{four_region_spectra}
\end{figure*}

\begin{figure*}
\centering
\includegraphics[width=.95\columnwidth,trim=0 5 0 10]{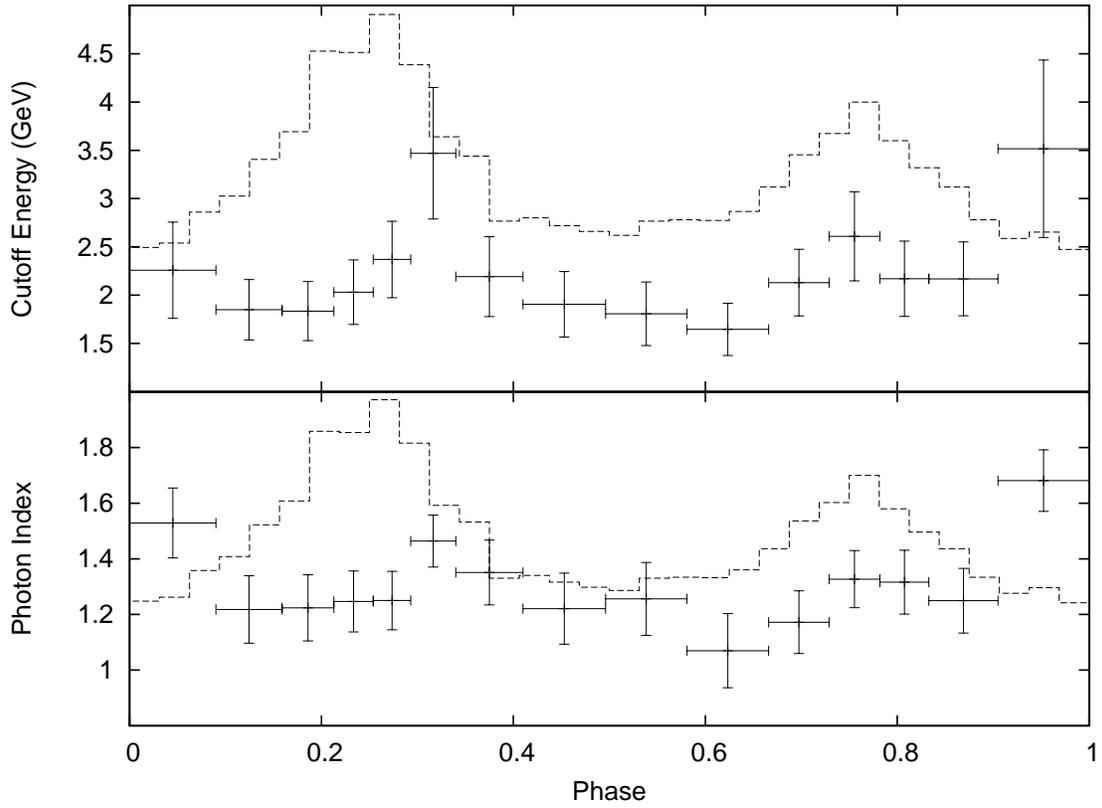}\\
\caption{Cutoff energy (top) and photon index (bottom) as a function of phase for PSR~J1836+5925 using 15 equal-count bins containing $\sim$650 events each. The dashed line in both panels shows the $>$100 MeV folded light curve of PSR~J1836+5925.}
\label{fig4}
\end{figure*}

Next, we studied the energy spectrum in the various phase intervals defined in Section~\ref{lc}. The spectral parameters of the pulsar were allowed 
to be free, while those of the other sources were fixed to the values obtained in the phase-averaged analysis. 
Figure~\ref{four_region_spectra} shows the four spectra and Table~\ref{spectral_results} summarizes the results, normalized for the different phase 
intervals, pointing to mild variations of the photon index $\Gamma$ and cutoff energy $E_{\rm cutoff}$ over the four phase intervals, with the off-peak region 
characterized by a softer spectrum. We also carried out a phase-resolved spectral analysis by taking 15 equal-counts bins of $\sim$650 photons. 
As in the previous analysis, the parameters of all the sources in the ROI were fixed at the values obtained in the phase-averaged analysis. 
Figure~\ref{fig4} shows the results of our analysis. The top panel illustrates the evolution of the cutoff energy, while the bottom panel shows the 
change in photon index with phase. Insufficient photon statistics prevent us from investigating the apparent spectral changes within the peaks in any 
finer detail.

\subsection{Variability analysis}

\subsubsection{Pulse profile variability}
We checked for variability in the pulse profile shape using non parametric tests. First, the data were divided into time segments with equal counts (from 2 
up to 32 segments). For every pair of segments, light curves were compared against the null hypothesis that both were drawn from the same parent 
distribution by performing Kolmogorov-Smirnov tests. There was no instance in which the pairwise comparison resulted in the null hypothesis being rejected 
even at the 80\% confidence level. $\chi^2$ tests yielded the same results, resulting in an overall gaussian distribution for the normalized residuals. We 
repeated the tests for light curves with 8, 10, 12, 15, 20, and 30 bins, confirming in every case that there is no hint of variability on timescales longer 
than a week.

\subsubsection{Flux variability}
\begin{figure*}
\centering
\includegraphics[width=.85\columnwidth,trim=0 5 0 10]{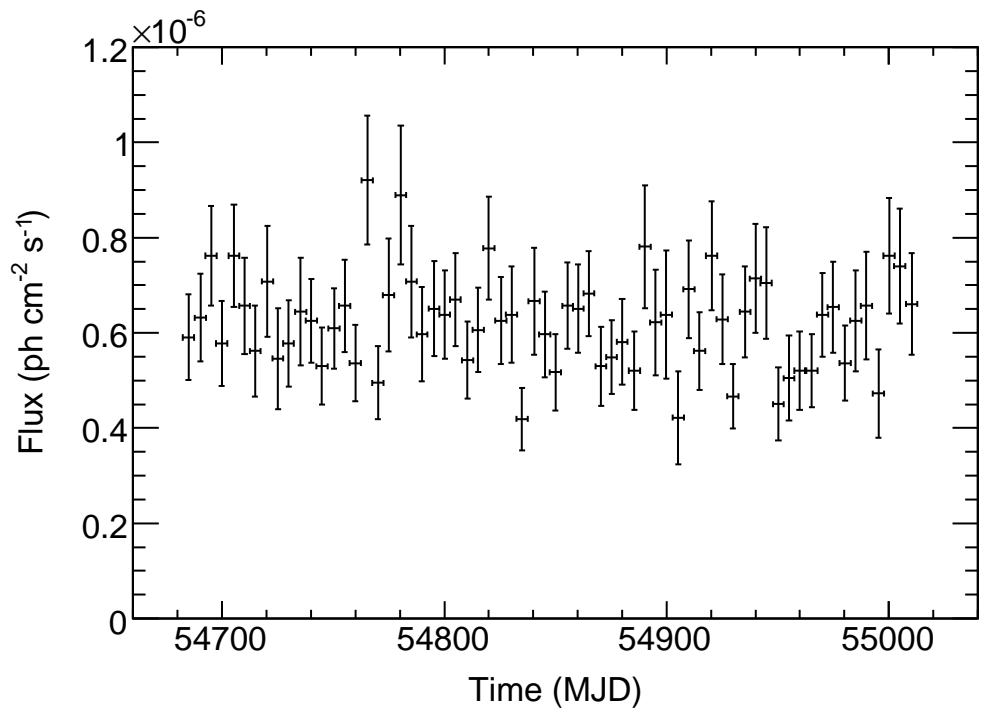}\\
\caption{Flux of PSR~J1836+5925 as a function of time in 5-day time bins, showing no evidence for variability (see Section 2.4.2).}
\label{flux}
\end{figure*}

In order to check the long term stability of the source, we computed the 0.1--100\,GeV photon flux in 5-day time bins using the LAT Science Tool 
\texttt{gtlike}. First, we constructed a spectral model file including all the LAT first-year catalog point sources catalog~\citep{LATCatalog} in our ROI, the 
Galactic diffuse emission and the isotropic background. Since AGN in the region of interest can be variable, the spectral parameters of all point sources were set free while 
the Galactic and isotropic background were fixed to the values obtained in the first phase-averaged analysis. After running \texttt{gtlike}, we checked the fit result for 
sources with large uncertainties. After removing these sources with low significance ($< 2\sigma$), we ran \texttt{gtlike} again to obtain the flux of 
PSR~J1836+5925 for that 5-day bin. Figure~\ref{flux} shows the resulting fluxes and statistical uncertainties. 
We used five-day bins as this was the interval chosen by \cite{bulgarelli08}, facilitating the comparison with the AGILE 
results. Assuming a constant flux, a $\chi^2$ test gives 
a value of 66.2 with 65 degrees of freedom. The variability index of \cite{mclaughlin96} is $V = 0.37$, consistent with no variation. The 
weighted standard deviation of the flux is 16\% of the average. If the flux is assumed to change linearly with time, the slope is consistent 
with zero, with a 68\% limit of 6.8\% change from beginning to end of the data. Fitting a sinusoidal variation produces an amplitude 
of $(2.5 \pm 3.9)$\% of the mean flux, consistent with zero.
We note that while the overall flux seen by the LAT agrees with that reported by the EGRET experiment~\citep{3EGcatalog}, it is somewhat higher than what has been reported 
by the AGILE experiment~\citep{bulgarelli08,AGILEcatalog}.

\section{Multi-wavelength Analysis}

\subsection{Radio search}

We used the LAT ephemeris to search anew for radio pulsations from PSR~J1836+5925. We folded the 24\,hr dataset obtained at the
GBT in 2002 December \citep[for details, see][]{halpern07} modulo the predicted period, while searching in dispersion
measure up to $\mbox{DM} = 100$\,cm$^{-3}$\,pc, using PRESTO~\citep{RansomThesis}. For the distance range 250--800\,pc~\citep{halpern07}, 
the DM predicted by the NE2001 electron density model~\citep{cordeslazio} is 2--9\,cm$^{-3}$\,pc. Although there is no evidence in 
the LAT 2008--2009 timing solution for rotational instabilities, and we do not expect a large degree of timing noise from such a 
relatively old pulsar, we also did a small search in period about the nominal value. No radio pulsations were detected.
For an assumed pulsar duty cycle of 10\%, our long observation at a frequency of 0.8\,GHz yields a flux density of $S_{0.8} < 7\,\mu$Jy (this is a significant improvement over the limit presented in \cite{halpern07} for the same data because we are now searching for a known period, allowing for a lower signal-to-noise ratio detection threshold).
Converted to the more usual pulsar search frequency of 1.4\,GHz with a typical spectral index of --1.6, $S_{1.4} <
3\,\mu$Jy.  The implied luminosity is $L_{1.4} \equiv S_{1.4} d^2 < 0.002 d_{0.8}^2$\,mJy\,kpc$^2$. This is at least an order of 
magnitude smaller than the least luminous radio pulsar, PSR~J1741$-$2054, originally discovered in gamma rays by the LAT 
\citep[see][]{camilo09}, suggesting that if PSR~J1836+5925 is an active radio pulsar its beam probably does not intersect the Earth. One 
caveat to this conclusion is that scintillation caused by the interstellar medium could be quite significant for this observation of this pulsar: 
depending on its actual DM within the expected range, the characteristic scintillation bandwidth and timescale at 0.8\,GHz predicted by the NE2001 
model may be greater than the observation bandwidth and time. If so, the received flux density of the pulsar during the observation
may not reflect its intrinsic average. To address this potential concern, we did one extra observation
with the GBT.  On 2009 October 24 we recorded data from a
bandwidth of 100\,MHz centered on 350\,MHz for 2.0\,hr using
GUPPI\footnote{https://wikio.nrao.edu/bin/view/CICADA/GUPPiUsersGuide}.
Again, no pulsations were detected from PSR~J1836+5925. For the same
assumed duty cycle, the flux limit was $55\,\mu$Jy. With the same
assumed spectral index, this corresponds to $14\,\mu$Jy at 0.8\,GHz.
While this recent 350\,MHz observation was thus nominally only half as
sensitive as the earlier 24\,hr observation at 820\,MHz, it was still
an extremely deep observation, and much more immune to scintillation
effects, rendering our earlier conclusion valid: for all practical
purposes, PSR~J1836+5925 is a ``radio quiet'' pulsar.

\subsection{X-ray observations\label{xray_section}}

We studied the X-ray counterpart of PSR~J1836+5925 using two \xmm observations taken on 
2008 May 18 and 2008 June 25 (15\,ks each). Both the EPIC/pn~\citep{strueder01} and 
the EPIC/MOS~\citep{turner01} cameras were operated in their Full Frame mode, using 
the thin optical filter. We focused on spectroscopy (the time resolution is not adequate 
to search for pulsations). Data reduction and analysis were performed with the \xmm 
Science Analysis Software (SASv8.0). 
Owing to the lack of variability between the two epochs, the 
two data sets were merged, resulting in 24.4\,ks, 30.7\,ks and 30.8\,ks of good 
exposure in the pn, MOS1 and MOS2 cameras, respectively. The background-subtracted 0.2--3 keV count rate 
of the source, as extracted from a 15\arcsec\ radius circle, is $0.023\pm0.001$\,cts s$^{-1}$, 
$0.0043\pm0.0004$\,cts s$^{-1}$, and $0.0040\pm0.0004$\,cts s$^{-1}$ in the pn, MOS1 
and MOS2 cameras, respectively. Background accounts for $\sim30\%$ additional counts.

We performed simultaneous fits to
the pn, MOS1 and MOS2 spectra using the XSPEC v12.4 software. The
X-ray spectrum cannot be described by a pure blackbody model
($\chi^2_{\nu}=3.70$, 40 d.o.f.). A simple power law is possibly
consistent with the data ($\chi^2_{\nu}=1.39$, 40 d.o.f.),
however, the best fit requires a rather large photon index
($\Gamma=3.0\pm0.2$) and a very low $N_H$ of $<2\times 10^{19}$
cm$^{-2}$ (errors are at $90\%$ confidence level for a single
parameter).

The combination of a blackbody and a power law yields a better
fit ($\chi^2_{\nu}=0.71$, 38 d.o.f.). The best fit model features
an absorbing column $N_H<2.7\times10^{20}$ cm$^{-2}$ (the best
fit value is 0), a blackbody temperature\footnote{We quote blackbody temperatures and emitting
  radii as measured by a distant observer throughout the paper.} kT = $59_{-17}^{+7}$
eV, and a power law photon index $\Gamma=1.7\pm0.3$. The total observed flux in 0.2-5 keV is $5.5\times10^{-14}$ erg
cm$^{-2}$ s$^{-1}$. Table~\ref{mw_results} summarizes the results of our best spectral fit. 

While our flux values are not dissimilar from those of \cite{halpern02}, the black-body temperature, 
as well as the $N_H$ values can now be constrained directly on the basis of the $XMM$ data.

The rather high value of the blackbody temperature, coupled with the
very low $N_H$ (consistent with 0) point to a small emitting surface
at a relatively low distance. For the best fitting temperature, an
emitting surface of 1 km radius would imply a 450\,pc distance
which would scale to 300\,pc for a 50 eV temperature. In no way
can emission from the entire neutron star be invoked since it
would imply a distance in excess of 3 kpc, not compatible with
our very low $N_H$. In such a scenario the contribution of the thermal emission from the
bulk of the surface should be negligible within the EPIC band. This
requires\footnote{In order to get a rough estimate, we fixed all spectral
parameters of the blackbody plus powerlaw model to their best fitting values,
leaving $N_H$ as a free parameter (with a maximum allowed value of $2.7 \times
10^{20}$ cm$^{-2}$), and we added a second blackbody component 
to account for surface emission.} a surface temperature
lower than $\sim25$ eV ($\sim30$ eV), assuming a 10 km emitting radius at 300 pc (450 pc),
which is consistent with expectations for a $\sim10^6$ yr old neutron star. 
Such a cooler thermal component would also be consistent with the deep HST upper limits.

\begin{figure*}
\centering
\includegraphics[width=.85\columnwidth,angle=-90,trim=0 5 0 10]{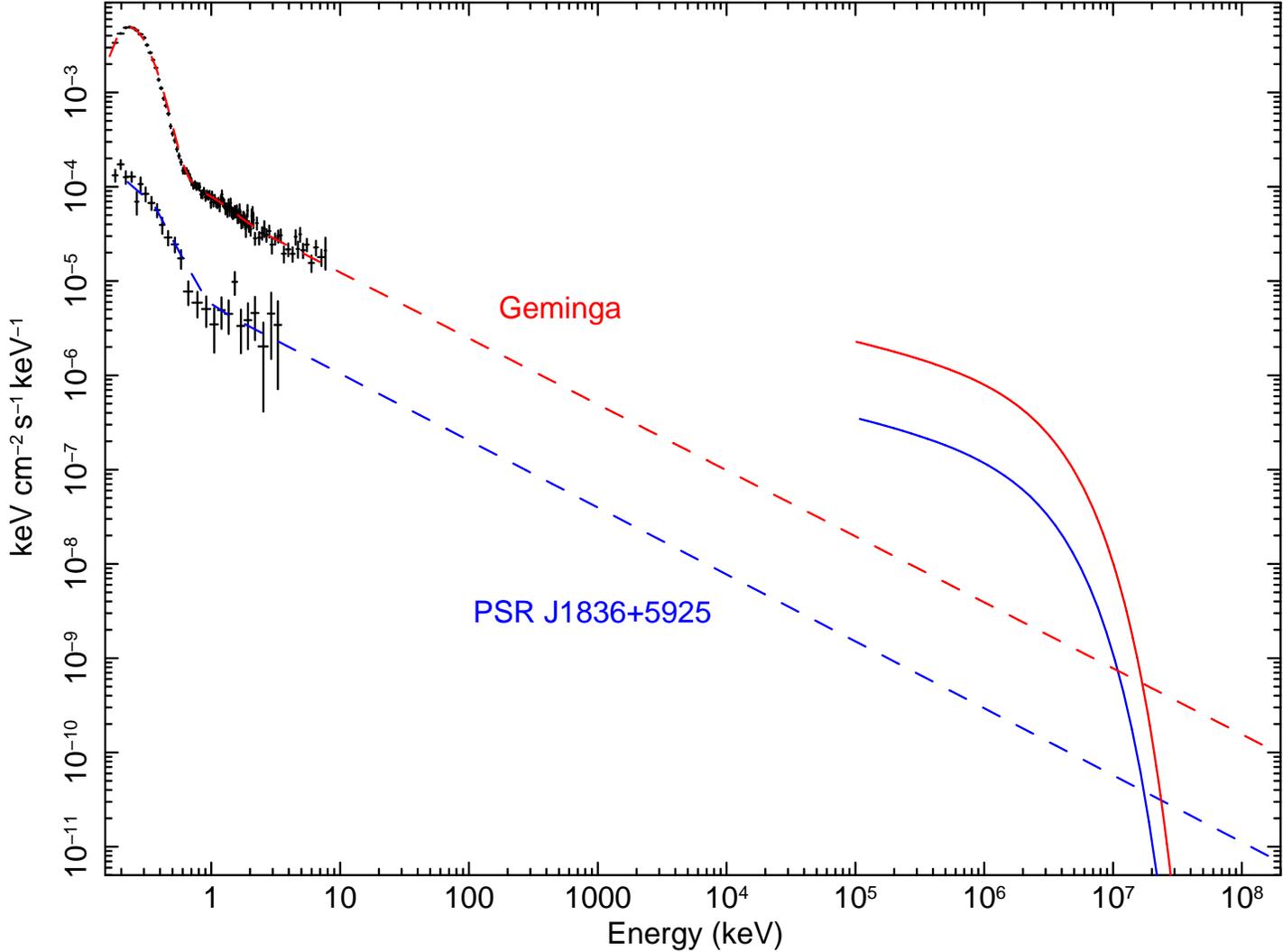}\\
\caption{Unfolded \xmm spectrum of PSR~J1836+5925, compared to that of Geminga, whose data have been reprocessed to 
take advantage of the new calibration files which cover the energy range down to 0.15 keV (EPIC status of
calibration and data analysis Document XMM-SOC-CAL-TN-0018 http://xmm2.esac.esa.int/docs/documents/CAL-TN-0018.pdf). The best fit models are 
superimposed \citep[for a thorough report on Geminga, see][]{caraveo03}. We extrapolate the models out to the LAT energy range and show the 
best-fit phase-averaged LAT spectra for the two pulsars. The overall similarity is apparent.}
\label{xrayfig}
\end{figure*}

The overall similarity of the X-ray spectrum of PSR~J1836+5925 and
Geminga is apparent in Figure~\ref{xrayfig}, where both the $XMM$ spectra and curves
fitting the phase-averaged LAT spectra for both these sources are
depicted.  This similarity applies to both the X-ray and $\gamma$-ray
spectra individually, to the LAT-band turnovers, to the offset between
their fluxes in each waveband, and therefore to the overall
multiwavelength impression~\citep[for the detailed LAT results on Geminga, see][]{LATGeminga}. This broadband 
picture clearly illustrates that PSR~J1836+5925 resembles Geminga in its high energy components, a
character that will guide future spectral modeling. The extrapolation
of the $XMM$ power law tails up to the LAT band highlights the disparity
between the X-ray non-thermal indices $\Gamma\sim 1.7$ and the LAT band
power law indices $\Gamma\sim 1.3$.  This property suggests that some as
yet undetectable spectral structure or feature must exist in the 20 keV
-- 100 MeV band, perhaps due to a transition between components spawned
by different radiative processes. The structure may be a simple
flattening, or something more complex, however it should be unlike
the steepening seen in the broadband X-ray/$\gamma$-ray spectrum of the 
younger Vela pulsar~\citep{strickman99}. The diagnostic potential enabled by the 
detection of such spectral structure motivates the development of future 
sensitive spectroscopy telescopes in the hard X-ray and soft $\gamma$-ray bands.


We also re-analyzed archival {\it Chandra} data taken in high time resolution, to search for possible X-ray pulsations. 
Using 118 ks of HRC-S data \citep[the same dataset described by][]{halpern07}, we extracted $\sim790$ source counts. No significant 
modulation is apparent in a 10 bin phase histogram folding the events with the extrapolated LAT timing solution. Although we expect no 
significant timing noise from this pulsar, a search for pulsations was also performed around a narrow range of the expected period (0.1732--0.1733 s), 
but no significant signal was detected. Following \cite{vaughan94}, we set an  upper limit of 40\% on the pulsed fraction (at $99\%$ confidence level), 
assuming a sinusoidal modulation. 

\section{Discussion}

The discovery by the LAT of PSR~J1836+5925 confirmed the long-held suspicion that 3EG~J1835+5918 was a nearby Geminga-like 
pulsar. Its characteristic age of 1.8 million years agrees with expectations that it should be significantly 
older than Geminga, given the soft X-ray spectrum of its X-ray counterpart, the absence of optical and radio emission, 
and the measured upper limits on its proper motion~\citep{halpern07}. We see no evidence for time 
variability of either the source flux or the pulse profile shape over the 11 months of observations. Our measured pulsations
explain why this pulsar proved rather difficult to detect: the small ($\sim$30\%) pulse fraction 
and relatively large duty cycle made blind searches of EGRET data futile. Indeed, it necessitated the LAT
pointed observations to get an early pulse detection. Our detection, in turn, raises several puzzles -- why does this object 
have such a large off-peak component and how does the relatively low spindown power produce an apparently large 
$\gamma$-ray luminosity for even modest distances?

        While other pulsars show detectable emission throughout most of the pulsar period~\citep[e.g. see][]{LATVela,LATCrab}, that 
of PSR~J1836+5925 is particularly bright, and provides a good opportunity to test the nature of these off-peak components. Since 
the source at pulse minimum is unresolved, and since the pulse minimum spectrum demands a 2--3\,GeV 
cutoff, we conclude that we are not seeing evidence for a surrounding pulsar wind nebula (PWN). The lack of 
extended emission in the deep {\it Chandra} images also implies that there is no bright PWN. 
Thus the off-peak flux is likely magnetospheric. Since the $\Gamma\approx 1.6$ off-peak spectral index is substantially 
softer than that of the rest of the profile, we tested whether a second, spatially unresolved 
pulsar could contribute the off-peak flux. We searched for pulsations from the source by applying the standard time-differencing technique~\citep{atwood06}, 
masking the frequency of PSR~J1836+5925. We used a maximum frequency of 64 Hz, and a long time-difference window of $\sim$12 days. We found no 
evidence for pulsations at any other frequency. The precise sensitivity of the blind search is still not completely understood, however, a comparison of
the blind search pulsars discovered so far~\citep{BSPI} and the known radio pulsars detected by the LAT,
suggests that the blind search is approximately 2--3 times less sensitive than a standard pulsation search using the known timing 
solution~\citep{pulsarcatalog}. This results in a 5$\sigma$ limit on the pulsed flux of $\sim2\times10^{-7}$ cm$^{-2}$ s$^{-1}$ for another putative 
pulsar at this location. We conclude that PSR~J1836+5925 emits over half its flux in a nearly constant component with an exponentially cut-off 
spectrum which is softer than the peaks of the profile.

The distance inferred from the observed $\gamma$-ray flux,
$F_\gamma = 6 \times 10^{-10}\, {\rm erg\,cm^{-2}\,s^{-1}}$, depends on the intrinsic luminosity ($L_\gamma$), the beam geometry, and the
line of sight along which we sample the anisotropic emission. To account for anisotropy, we parameterize the relation between the observed 
flux and the true luminosity by the ``flux conversion factor'' $f_\Omega = L_{\gamma}/4\pi d^2 F_{\gamma}$, whose estimation we discuss in the next 
paragraph. The intrinsic luminosity can be inferred from the spindown luminosity of the pulsar, if we know the efficiency $\eta = L_\gamma / \dot E$. We 
therefore have $d = ({\dot E}/4\pi F_{\gamma})^{1/2}(\eta/f_\Omega)^{1/2}$. It has been argued that the efficiency of $\gamma$-ray emission, and the fraction 
of the open zone participating in the gaps, grows with decreasing ${\dot E}$~\citep{ruderman88,arons96}, and observations support 
$\eta \propto {\dot E}^{-1/2}$~\citep{pulsarcatalog}. We adopt $\eta = C\,({\dot E} /  10^{33}{\rm erg\,s^{-1}})^{-1/2}$, where $C$ is a slowly varying function of 
order unity which depends on the details of the physical model~\citep{watters09}. The observed $\dot E=1.1\times10^{34}$\,erg s$^{-1}$ then implies an 
efficiency of $\eta\sim0.30$. Using this efficiency, along with the known values for $\gamma$-ray flux $F_\gamma$ and spindown luminosity ${\dot E}$, our 
estimate for the pulsar distance becomes $d \approx 215 f_\Omega^{-1/2}$\,pc.

The factor $f_\Omega$ depends sensitively on the emission model, on the inclination
of the pulsar spin axis to the line of sight ($\zeta$), and on the inclination 
of the magnetic pole with respect to the spin axis ($\alpha$). Models are described in
\cite{watters09}; for the ``Two Pole Caustic" (TPC) model, pulse separations $\Delta = 0.5$ occur in 
two regions: $\alpha \ga 85^\circ$, $\zeta \la 60^\circ$ or $\alpha \la 60^\circ$, $\zeta \ga 85^\circ$~\citep[near the axes in the magenta 
zone of Figure 3 in][]{watters09}. The 
former solutions are, however, not satisfactory as they have weak bridge fluxes, at least for models with thin 
radiating surfaces. Thicker emission zones can produce additional bridge flux~\citep{venter09}. The large $\zeta$ 
solutions can indeed have substantial off-peak flux arising at modest $r<0.2\,r_{LC}$ altitudes (where $r_{LC}=cP/2\pi$ is the 
speed of light cylinder), especially for efficiencies $\eta \la 0.2$. For the outer gap (OG) model, only a few $\zeta \ga 80^\circ$, 
$\alpha \la 30^\circ$ models give the observed $\Delta$ for highly efficient pulsars ($\eta \sim 0.2$). These have relatively 
large off-peak fluxes, arising from large $r>0.5\, r_{LC}$ altitudes.

        While both models have acceptable large $\zeta$ solutions, for the TPC model the 
$f_\Omega$ is typically $0.9\pm 0.1$, with the small $\alpha$ solutions trending
to $f_\Omega >2$. In contrast, the few acceptable OG models have $f_\Omega \la 0.1$.
The resulting distance for the TPC model is typically 
$d\approx 250$\,pc (but in some cases can be $d \la 170$\,pc). For the OG model we expect $d \approx 750$\,pc.
Both cases are small enough to be compatible with the small X-ray absorption discussed in Section~\ref{xray_section}.
In summary, the TPC model has more acceptable solutions, but would imply a very small distance. The relatively small parameter space
of acceptable OG solutions, on the other hand, is offset by a larger source distance and hence a larger Galaxy volume in which such a pulsar could be found. 

	In general, the results of our X-ray analyses are in broad agreement with previous investigations~\citep{halpern02}, which 
were based on a factor $>3$ smaller photon statistics. Our analysis clearly shows that the spectrum of the candidate counterpart is indeed 
consistent with the one of a nearby, thermally-emitting, middle-aged isolated neutron star.

        The best prospect for refining our understanding of the emission from PSR~J1836+5925 would clearly come from an 
accurate distance measurement. This seems difficult to obtain, although improved X-ray spectral 
measurements and models could help. Alternatively, if $\gamma$-ray pulsar spectral models can be developed 
sufficiently, we may be able to connect the softer off-peak spectrum with a particular magnetospheric
location. In either case, the low power, large characteristic age and relatively close distance 
imply that J1836+5925 is the harbinger of a large population of old and weak $\gamma$-ray pulsars~\citep{BSPI,BSPII}.

\begin{acknowledgments}

The \textit{Fermi} LAT Collaboration acknowledges generous ongoing support
from a number of agencies and institutes that have supported both the
development and the operation of the LAT as well as scientific data analysis.
These include the National Aeronautics and Space Administration and the
Department of Energy in the United States, the Commissariat \`a l'Energie Atomique
and the Centre National de la Recherche Scientifique / Institut National de Physique
Nucl\'eaire et de Physique des Particules in France, the Agenzia Spaziale Italiana
and the Istituto Nazionale di Fisica Nucleare in Italy, the Ministry of Education,
Culture, Sports, Science and Technology (MEXT), High Energy Accelerator Research
Organization (KEK) and Japan Aerospace Exploration Agency (JAXA) in Japan, and
the K.~A.~Wallenberg Foundation, the Swedish Research Council and the
Swedish National Space Board in Sweden.

Additional support for science analysis during the operations phase is gratefully
acknowledged from the Istituto Nazionale di Astrofisica in Italy and the Centre National d'\'Etudes 
Spatiales in France.

The GBT is operated by the National Radio Astronomy Observatory, a
facility of the National Science Foundation operated under cooperative
agreement by Associated Universities, Inc.

This work is partly based on observations obtained with $XMM-Newton$, an ESA science mission with instruments and 
contributions directly funded by ESA Member States and NASA.

\end{acknowledgments}

\bibliographystyle{apj}
\bibliography{journapj,Fermi_Bibtex_full_v2,my_bibliography}

\begin{table*}[p]
\centering
\caption{Measured and derived timing parameters for PSR\,J1836+5925 \label{timing_results}} {\footnotesize
\begin{tabular}{ll}
\hline
Parameter 				& Value~\tablenotemark{a}\\ 
\hline
MJD range			& 54647.4--55013.0 \\
Epoch (MJD)			& 54800 \\
R.A. (J2000) 			&  18:36:13.75(3) \\
Dec. (J2000)			& +59:25:30.3(6) \\
$\nu$ (Hz) 			&  5.7715514983(4) \\
$\dot{\nu}$ (Hz s$^{-1}$) 	&  --4.97(2)$\times10^{-14}$ \\
Rms timing residual (ms)	& 1.3 \\
Characteristic age, $\tau_c$ (kyr) & 1840 \\
$\dot E  ~($erg s$^{-1})$	&  1.1$\times10^{34}$\\
Surface magnetic dipole field strength (gauss)		& 5.1$\times10^{11}$ \\
$\gamma$-ray peak separation ($\Delta$)	& $0.51\pm0.01$	\\
\hline 
\end{tabular}
\tablenotetext{a}{The numbers in parentheses are the 1$\sigma$ uncertainties
derived from the timing model (see Section 2.1).}
\label{results_table}
}
\end{table*}
\newpage

\begin{table*}[p]
\centering
\caption{LAT $\gamma$-ray spectral results PSR\,J1836+5925\label{spectral_results}} {\footnotesize
\begin{tabular}{lll}
\hline
Phase region~\tablenotemark{a} & Parameter 				& Value~\tablenotemark{b}\\ 
\hline
 		& MJD range 			&  54682.7--55013.0 \\
		& 				&			\\	
Phase-averaged		& Photon Flux, F$_{100}$ (ph cm$^{-2}$ s$^{-1}$)	&	$(6.24\pm0.12)\times10^{-7}$ 	\\
$0 < \phi < 1$		& Energy Flux, G$_{100}$ (erg cm$^{-2}$ s$^{-1}$)	&	$(5.91\pm0.08)\times10^{-10}$	\\
			& Photon Index, $\Gamma$				&	$1.31\pm0.03$	\\
			& Cutoff Energy, $E_{\rm cutoff}$ (GeV)			&	$2.27\pm0.11$	\\
			& Normalization, $K$ (ph cm$^{-2}$ s$^{-1}$ MeV$^{-1}$)	&	$(1.82\pm0.06)\times10^{-10}$	\\
			& 				&			\\
First Peak (FP)		& Photon Flux, F$_{100}$ (ph cm$^{-2}$ s$^{-1}$)	&	$(8.32\pm0.24)\times10^{-7}$	\\
$0.105 < \phi < 0.405$	& Energy Flux, G$_{100}$ (erg cm$^{-2}$ s$^{-1}$)	&	$(7.97\pm0.17)\times10^{-10}$	\\
			& Photon Index, $\Gamma$				&	$1.31\pm0.05$		\\
			& Cutoff Energy, $E_{\rm cutoff}$ (GeV)			&	$2.31\pm0.17$		\\
			& Normalization, $K$ (ph cm$^{-2}$ s$^{-1}$ MeV$^{-1}$)	&	$(2.42\pm0.12)\times10^{-10}$	\\			
			& 				&			\\
Second Peak (SP)	& Photon Flux, F$_{100}$ (ph cm$^{-2}$ s$^{-1}$)	&	$(6.44\pm0.22)\times10^{-7}$	\\
$0.632 < \phi < 0.904$	& Energy Flux, G$_{100}$ (erg cm$^{-2}$ s$^{-1}$)	&	$(6.40\pm0.16)\times10^{-10}$	\\
			& Photon Index, $\Gamma$				&	$1.24\pm0.05$		\\
			& Cutoff Energy, $E_{\rm cutoff}$ (GeV)			&	$2.18\pm0.18$		\\
			& Normalization, $K$ (ph cm$^{-2}$ s$^{-1}$ MeV$^{-1}$)	&	$(2.03\pm0.12)\times10^{-10}$	\\			
			& 				&			\\
Bridge (BR)		& Photon Flux, F$_{100}$ (ph cm$^{-2}$ s$^{-1}$)	&	$(4.47\pm0.28)\times10^{-7}$	\\
$0.459 < \phi < 0.597$	& Energy Flux, G$_{100}$ (erg cm$^{-2}$ s$^{-1}$)	&	$(4.09\pm0.17)\times10^{-10}$	\\
			& Photon Index, $\Gamma$				&	$1.17\pm0.11$		\\
			& Cutoff Energy, $E_{\rm cutoff}$ (GeV)			&	$1.64\pm0.23$		\\
			& Normalization, $K$ (ph cm$^{-2}$ s$^{-1}$ MeV$^{-1}$)	&	$(1.66\pm0.20)\times10^{-10}$	\\			
			& 				&			\\
Off-Peak (OP)		& Photon Flux, F$_{100}$ (ph cm$^{-2}$ s$^{-1}$)	&	$(4.87\pm0.38)\times10^{-7}$	\\
$0.053 > \phi > 0.938$	& Energy Flux, G$_{100}$ (erg cm$^{-2}$ s$^{-1}$)	&	$(3.63\pm0.19)\times10^{-10}$	\\
			& Photon Index, $\Gamma$				&	$1.59\pm0.11$		\\
			& Cutoff Energy, $E_{\rm cutoff}$ (GeV)			&	$2.65\pm0.57$		\\
			& Normalization, $K$ (ph cm$^{-2}$ s$^{-1}$ MeV$^{-1}$)	&	$(9.90\pm1.2)\times10^{-11}$	\\
\hline
\end{tabular}
\tablenotetext{a}{See the top panel in Figure 1 for a visual representation of the various phase regions.}
\tablenotetext{b}{All errors quoted are statistical. In addition, systematic errors of $\delta$F$_{100}$ = (+30\%, --10\%), $\delta$G$_{100}$ = (+20\%, --10\%), 
$\delta\Gamma$ = (+0.3, --0.1), and $\delta E_{\rm cutoff}$ = (+20\%, --10\%) must be taken into account~\citep{pulsarcatalog}.}

\label{results_table}
}
\end{table*}
\newpage

\begin{table*}[p]
\centering
\caption{Radio and X-ray results for PSR\,J1836+5925\label{mw_results}} {\footnotesize
\begin{tabular}{lll}
\hline
Wavelength & Parameter 				& Value \\ 
\hline
\bf{Radio} 		& Radio flux density at 350 MHz, $S_{0.35} (\mu$Jy) 	&  $< 55$		\\
	 		& Radio flux density at 0.8 GHz, $S_{0.8} (\mu$Jy) 	&  $< 7$		\\
			& Radio flux density at 1.4 GHz, $S_{1.4} (\mu$Jy)~\tablenotemark{a}	&  $< 3$		\\				
\bf{X-ray}		& (0.2--5 keV) Total observed~\tablenotemark{b} X-ray flux (erg cm$^{-2}$ s$^{-1}$)	&  5.5$\times10^{-14}$	\\
			& (0.2--5 keV) Unabsorbed non-thermal flux (erg cm$^{-2}$ s$^{-1}$)	&  3.0$\times10^{-14}$	\\
			& X-ray blackbody temperature, kT (eV)			&  $59_{-17}^{+7}$	\\
			& X-ray blackbody radius (km)				&  ($1.5_{-0.4}^{+5.3}$)d$_{0.8}$~\tablenotemark{c}	\\
			& X-ray absorbing column, $N_H$ (cm$^{-2}$)		&  $< 2.7\times10^{20}$ \\
			& X-ray power law photon index, $\Gamma$		&  1.7$\pm$0.3	\\
			& X-ray power law normalization at 1 keV, N$_{\mathrm{PL}}$ (ph cm$^{-2}$ s$^{-1}$ keV$^{-1}$)	& (5.7$\pm1.0)\times10^{-6}$	\\  
\hline 
\end{tabular}
\tablenotetext{a}{There is no measured upper limit at 1.4 GHz. This limit is derived from the observation made at 0.8 GHz (See Section 3.1)}
\tablenotetext{b}{Total observed and unabsorbed flux coincide since the best fitting column density is $N_H$=0.}
\tablenotetext{c}{d$_{0.8}$ is the distance to PSR\,J1836+5925 in units of 0.8 kpc.}
\label{results_table}
}
\end{table*}
\newpage

\newpage

\newpage

\newpage

\newpage

\newpage

\end{document}